# Structure of Polypeptide Based Diblock Copolymers in Solution: Stimuli-responsive Vesicles and Micelles

*Frédéric Chécot[1], Annie Brûlet[2], Julian Oberdisse[3], Yves Gnanou[1], Olivier Mondain-Monval[4], and Sébastien Lecommandoux[1] ∗*

[1]Laboratoire de Chimie des Polymères Organiques (LCPO-UMR5629),

ENSCPB – Université Bordeaux I

16, Av. Pey Berland, 33607 Pessac Cedex (France)

[2] Laboratoire Léon Brillouin (CEA-CNRS UMR12), CEA-Saclay

91191 Gif sur Yvette Cedex (France)

[3] Laboratoire des Colloïdes, Verres et Nanomatériaux (LCVN),

Université Montpellier II, Place E. Bataillon,

34095 Montpellier Cedex 05 (France)

[4] Centre de Recherche Paul Pascal (CNRS – University Bordeaux 1)

Av. Schweitzer, 33600 Pessac Cedex (France)

* Corresponding author : Sébastien Lecommandoux. E-mail: s.lecommandoux@enscpb.fr

**Abstract**

Polypeptide-based diblock copolymers forming either well-defined self-assembled micelles or vesicles after direct dissolution in water or in dichloromethane, have been studied combining light and neutron scattering with electron microscopy experiments. The size of these structures could be reversibly manipulated as a function of environmental changes such as pH and ionic strength in water. Compared to other pH-responsive seld-assembled systems based on "classical" polyelectrolytes, these polypeptide-based nanostructures present the ability to give a response in highly salted media as the chain conformational ordering can be controlled. This makes these micelles and vesicles suitable for biological applications: they provide significant advantages in the control of the structure and function of supramolecular self-assemblies.

**I-   Introduction**

One of the greatest challenges facing physics, chemistry, and materials science today concerns the design of molecules that either spontaneously or upon application of a trigger can build functional superlattices by self-organization. In order to obtain such ordered structures it is essential that both long range repulsive and short range attractive forces come in to play at the same time. In this way diblock copolymers consisting of two chemically different and flexible segments (coil-coil diblock copolymers) are well-known for their ability to self-organize in a variety of morphologies. The phase behavior of coil-coil diblock copolymers is well understood, both experimentally and theoretically.[1] Replacing one of the blocks of a coil-coil diblock copolymer by a rigid segment significantly changes the phase behavior. The self-organization of such rod-coil diblock copolymers is no longer solely induced by the phase separation of the constituent blocks, but is also influenced by the tendency of the rigid segments to aggregate. This dissymmetry between the flexibility of one block and the rigidity of the other can lead to morphologies that are distinctly different from those commonly observed in classical coil-coil diblock copolymers in bulk[2], and also in solution as shown recently with the formation of vesicles.[3,4]

Over the past few years, increasing attention has been given to the supramolecular organization of water-soluble block copolymer surfactants and to their potential use in applications such as coatings, drug delivery systems, nanoparticles or nanoreactors.[5] An important issue towards making these self-assembled systems useful for specific applications is their capability to respond to external stimuli such as temperature and/or pH.[6,7] The few water-soluble stimuli-responsive systems investigated so far were essentially based on the electrostatic repulsion between charged polyelectrolyte blocks (typically poly(2-vinylpyridine) $P_2VP$ and polyacrylamide PAA), but such systems are not completely reversible owing to the formation of salt with each cycle of pH variation. Finally, another important issue in the design of a self-assembled material is its stability. In fact, most aggregates are only stable within a certain range of concentration, temperature or pH. A suitable approach to improve the stability of self-assembled structures is to cross-link one of the blocks.[8,9]

Recently, block copolymers comprised of polypeptide segments have been shown to provide significant advantages in controlling both the function and supramolecular structure of bio-inspired self-assemblies. Deming et al.[3] demonstrated the ability of copolypeptides to form stimuli-responsive vesicles by conformation-specific assembly. Our group also demonstrated very recently that zwitterionic copolypeptides can form reversible self-assembled vesicles as a function of pH.[4] Nevertheless, most of the polypeptide-based amphiphilic copolymers studied so far were formed by a water-soluble polypeptide segment and a hydrophobic synthetic block (polybutadiene,[10,11] polyisoprene,[12] poly(ε-caprolactone)[13] or a glycopolymer[14]). As demonstrated by us, copolymers combining polybutadiene-*b*-polypeptide are very attractive and the resultant self-assembled structures can be covalently cross-linked to form stimuli-responsive nano-objects.[15]

We report here on the self-organization of a series of polybutadiene-*b*-poly(L-glutamic acid) (PB-*b*-PGA) diblock copolymers in aqueous or organic solutions, combining a large range of experimental analysis in order to describe in great details the self-assembly properties of such systems. Our aim is to exploit the pH-sensitivity of polypeptides' secondary structure to manipulate the size and shape of the supramolecular structures formed by self-organization of these block copolymers in aqueous media. Stimuli-responsive nanoparticles have been reported before, and typically relied on pH, ion-strength or solvent induced changes in the swelling behavior.[6] In contrast to most other amphiphilic polyelectrolyte (block) copolymers, the use of polypeptide based diblock copolymers for the preparation of such structures is much more interesting. Indeed, the hydrophilic peptide block has the ability to fold into well-defined secondary structures such as α-helix, β-sheet or coil depending on environmental changes (temperature, pH, ionic strength). More importantly, the swelling behavior of nanoparticles composed of peptide-based block copolymers can be regulated by pH changes even in high ionic strength media as demonstrated previously by our group.[10] Furthermore, it can be precisely controlled by appropriate design of the amino acid sequence. The self-assembly in water and in organic solvents of these copolymers will be described especially by means of static and dynamic light scattering, small angle neutron scattering and electron microscopy. Then we will investigate the response to pH variations of the self-organized vesicles and micelles in water.

## II- Experimental section

### II-1- Materials and reagents

All reagents and solvents were purchased from commercial suppliers and were used as received unless otherwise noted. 1,3-Butadiene was successively stirred at −35 °C and cryodistilled from *sec*-BuLi. Tetrahydrofuran (THF) was refluxed over $CaH_2$, distilled, stored over Na/Benzophenone, cryodistilled over sodium mirror, and cryodistilled in high vacuum prior to use. 1-(3-Chloropropyl)-2,2,5,5-tetramethyl-1-aza-2,5-disilacyclopentane was prepared via a literature procedure.[16] *N,N'*-Dimethylformamide (DMF) was distilled from $CaH_2$ under reduce pressure and subsequently stored over molecular sieves (4 Å) under an argon atmosphere. The γ-benzyl-L-glutamate N-carboxyanhydride (Bn-Glu NCA) was prepared according to a literature procedure.[17] All polymerization reactions were performed in flame-dried glassware employing standard high vacuum techniques.

### II-2- Polymerization reactions

The synthesis of primary amine end-functionalized polybutadiene blocks and the preparation of polybutadiene-*b*-poly(glutamic acid) diblock copolymers have already been described in detail in a previous publication.[15] We simply recall here that the synthesis of PB-*b*-PGA copolymers starts with the anionic polymerization of butadiene in THF at -78°C using sec-butyllithium (sec-BuLi) as initiator. In order to increase the extent of amine end-functionalization, the oligobutadienyllithium chains were first treated with diphenylethylene prior to be quenched with 1-(3-chloropropyl)-2,2,5,5-tetramethyl-1-aza-2,5-disilacyclopentane. After acidification followed by aqueous workup, crude ω-amino oligobutadiene (PB-$NH_2$) could be obtained. The PB-$NH_2$ oligomers have been separated from the non-functionalized polybutadiene by column chromatography on silica with a mixture dicloromethane/methanol (9/1 v/v) with a yield of 80%. The pure primary amine end-functionalized oligomer was then used to initiate the ring opening oligomerization of γ-benzyl-L-glutamate N-carboxyanhydride (Bn-GluNCA) in DMF.[18] The length of the γ-benzyl-L-glutamate segment could be controlled through the molar ratio of Bn-GluNCA to ω-amino oligobutadiene initiator leading to the expected polybutadiene-b-poly(γ-benzyl-L-glutamate) block copolymers. After deprotection of the benzyl groups by basic hydrolysis, amphiphilic diblock copolymers $PB_x$-b-$PGA_y$ were obtained (x and y respectively correspond to the number average degrees of polymerization of polybutadiene and poly(glutamic acid) blocks). Their molar masses have been determined by means of $^1H$ NMR, and GPC. The characteristics of all the peptide based diblock copolymers $PB_x$-b-$PGA_y$ are summarized in Table 1.

**Table 1: Characteristics of the diblock copolymers $PB_{48}$-b-$PGA_y$**

| $PB_{48}$-b-$PGA_y$[a] | PGA block[b] | | $\overline{M_n}$ Copolymer (g.mol$^{-1}$) |
|---|---|---|---|
| | $\overline{M_n}$ (g.mol$^{-1}$) | $\overline{DP_n}$ | |
| $PB_{48}$-b-$PGA_{20}$ | 2600 | 20 | 5200 |
| $PB_{48}$-b-$PGA_{56}$ | 7200 | 56 | 9800 |
| $PB_{48}$-b-$PGA_{114}$ | 14700 | 114 | 17300 |
| $PB_{48}$-b-$PGA_{145}$ | 18700 | 145 | 21300 |

[a] the same polybutadiene has been used as precursor with $\overline{M_n}$ = 2600 g.mol$^{-1}$ and $\overline{DP_n}$ = 48 as determined by combination of GPC and $^1H$-NMR measurements
[b] determined by $^1H$-NMR measurements

### II-3- Sample preparation

All the diblock copolymers, except for $PB_{48}$-*b*-$PGA_{20}$, readily dissolved in water upon addition of one eq. NaOH. The critical aggregation concentration (c.a.c), which is a measure of the stability and of the physical properties of the aggregate formed, was determined using pyrene as a fluorescent probe.[10] All the amphiphilic diblocks presented a c.a.c. around $5.10^{-6}$ mol.$L^{-1}$ which is typical for copolymers. This low c.a.c value corresponds to weight concentrations of about 50mg.$L^{-1}$ and indicates that the PB-*b*-PGA diblock copolymer forms stable aggregates in water. Moreover, according to the time dependence of the hydrodynamic radius ($R_H$) as determined by dynamic light scattering (DLS), thermodynamic equilibrium is established after about 10 days of vigorous stirring at room temperature and after 3 days at 40°C. The non water-soluble diblock $PB_{48}$-*b*-$PGA_{20}$ was dissolved in selective solvents for polybutadiene (THF and dichoromethane). As followed by DLS, stable self-assembled structures were formed almost instantaneously. All the experiments were performed above the c.a.c. (concentration range between 0.5% and 8%).

### II-4- Measurements

$^1$H NMR spectra of the deprotected diblock copolymers were recorded at room temperature on a Bruker Avance 400 MHz spectrometer using the residual proton resonance of the deuterated solvent as the internal standard.

Dynamic light scattering (DLS) and Static light scattering (SLS) experiments were performed using ALV Laser Goniometer, which consists of 22 mW HeNe linear polarized laser with 632.8 nm wavelength and an ALV-5000/EPP Multiple Tau Digital Correlator with 125 ns initial sampling time. The samples were kept at constant temperature (25.0 °C) during all the experiments. The accessible scattering angle range is from 10° up to 150°. However, most of the dynamic measurements were done at 90°. The solutions were introduced into 10 mm diameter glass cells. The minimum sample volume required for the experiment was 1 ml. The data acquisition was done with the ALV-Correlator Control Software, and the counting time varied for each sample from 300 s up to 600 s. Millipore water was thoroughly filtered with 0.1 µm filters and directly employed for the preparation of the solutions.

Transmission Electron Microscopy (TEM) pictures were recorded on a JEOL JEM100S microscope working at 80 KV. The samples were sprayed onto formvar coated cupper grids using a homemade toll. Freeze-fracture pictures were obtained from modified samples: 1.25 g.l$^{-1}$ copolymer solutions (70/30 v/v water/glycerol) that were fractured in a BAF 300 Balzers apparatus. Glycerol was used as a cryoprotectant in order to avoid the formation of ice crystals that would disrupt the sample structure.

Circular Dichroïsm (CD) experiments were carried out on a JOBIN YVON CD6 Spex spectrometer (184-900nm).

Small angle neutron scattering was performed at the Laboratoire Léon Brillouin (CEA-CNRS, Saclay) on the PACE spectrometer. Three configurations were used varying the wavelength λ and the sample-to-detector distance D in order to cover a broad wave vector range (λ=5Å, D=1.07m; λ=17Å, D=4.57m; λ=6Å, D=2.87m). Absolute values of the scattering intensity, I(q) in cm$^{-1}$, were obtained[19] from the direct determination of the incident beam flux. The scattering curves have been analyzed by fitting theoretical spectra, which takes into account the polydispersity of the object (using a log-normal size distribution), convoluted for each configuration with the corresponding resolution function,[20] including the angular resolution arising from the circular apertures of the collimation, the detector cell size and the wavelength distribution (FWHM 10%) of the mechanical velocity selector.

**III-    Results and discussion**

   **III-1- Self-assembly properties in solution**

The DLS autocorrelation functions have been measured for all the block copolymers in solution, in water at pH=12, or in THF and dichloromethane for the $PB_{48}$-*b*-$PGA_{20}$ sample. In order to avoid the electrostatic interactions, the experiments were systematically performed in 1M NaCl, in a concentration range 0.25 g.l$^{-1}$ < c < 2.5 g.l$^{-1}$ (well above the CAC) and for angles between 40° and 130°. The DLS correlation function was analyzed with the cumulant method.[21] For all the diblock copolymers PB-*b*-PGA, the resultant self-assembled structures showed monomodal distribution with a translational diffusive mode. The hydrodynamic radius ($R_H$) could be calculated from the Stokes-Einstein equation. Figure 1 represents the variation of the hydrodynamic radius ($R_H$) as a function of the concentration for all the block copolymers studied. These results show that the size of the micelles remains almost constant for the whole range of concentration investigated. Static light scattering measurements were also performed using the same conditions. Upon drawing the Zimm plot of the scattering intensity, as shown in Figure 2 for $PB_{48}$-*b*-$PGA_{56}$ we could determine the radii of gyration of copolymers. The hydrodynamic radii ($R_H$) and the radii of gyration ($R_G$) are given in Table 2. The ratio $\rho = R_G / R_H$ is a characteristic that depends on the morphology of the aggregates formed.[22] ρ values close to one can be attributed to a vesicle geometry, whereas smaller values are generally encountered with spherical micelles (theoretically ρ=0.775). From the obtained values from DLS measurements in the presence of salt, one can conclude that $PB_{48}$-*b*-$PGA_{114}$ and $PB_{48}$-*b*-$PGA_{145}$ form spherical micelles in water and that $PB_{48}$-*b*-$PGA_{56}$ and $PB_{48}$-*b*-$PGA_{20}$ self-assemble into vesicles respectively in water and in THF or $CH_2Cl_2$.

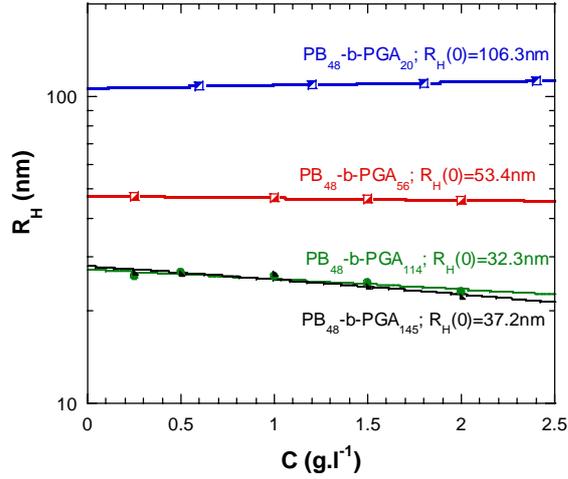 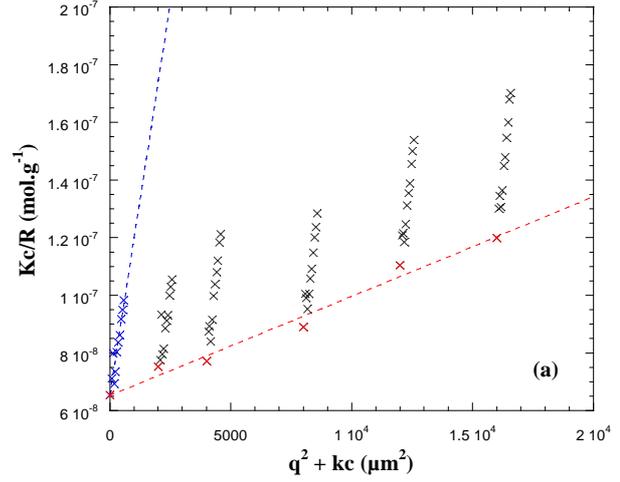

Figure 1: Evolution of $R_H$ as a function of the polymer concentration for $PB_{48}$-*b*-$PGA_{56}$, $PB_{48}$-*b*-$PGA_{114}$, and $PB_{48}$-*b*-$PGA_{145}$ in water (pH=12; 1M NaCl), and for $PB_{48}$-*b*-$PGA_{20}$ in dichloromethane.

Figure 2: Zimm plot of $PB_{48}$-*b*-$PGA_{56}$ in water (pH=12; 1M NaCl) obtained at 25°C over a concentration range of 0.25g/l to 2.5g/l. K is an apparatus constant ($K=4\pi n^2(dn/dc)^2/\lambda^4 N_{av}$), k=8000 is an arbitrary constant, q is the wave-vector ($q=\sin(\theta/2)4\pi/\lambda$) and c the polymer concentration.

TEM confirmed the global sizes and shapes of the self-assembled structures (the results are listed in Table 2). Indeed, as shown in Figure 3 copolymers $PB_{48}$-*b*-$PGA_{114}$ and $PB_{48}$-*b*-$PGA_{145}$ clearly show uniform size with a spherical shape corresponding to hard spheres. The radii measured on the pictures are in good agreement with the one determined from light scattering analysis. However, the TEM image corresponding to a freeze-fracture of copolymer $PB_{48}$-*b*-$PGA_{56}$ looks completely different and is compatible with the presence of a hollow structure for which the membrane is visible on the TEM picture (fig 3c).

**Table 2: DLS, SLS, SANS and TEM data obtained for the diblock copolymers PB-*b*-PGA.**

| | | | $PB_{48}$-*b*-$PGA_{56}$ | $PB_{48}$-*b*-$PGA_{114}$ | $PB_{48}$-*b*-$PGA_{145}$ | $PB_{48}$-*b*-$PGA_{20}$ | $PB_{48}$-*b*-$PGA_{20}$ |
|---|---|---|---|---|---|---|---|
| | mol% PGA | | 54 | 70 | 75 | 30 | 30 |
| | Solvent | | $H_2O$ | $H_2O$ | $H_2O$ | THF | $CH_2Cl_2$ |
| DLS | $R_{H(0\ NaCl)}$ | (nm) | 53 | 32 | 37 | 108 | 106 |
| | $R_{H(1M\ NaCl)}$ | | 47 | 27 | 28 | -- | -- |
| SLS | $R_{G(1M\ NaCl)}$ | (nm) | 50 | 19.5 | 23 | 111 | 103 |
| | $\rho = R_G / R_H$ | | 1.05 | 0.72 | 0.82 | 1.03 | 0.97 |
| | morphology | | *vesicle* | *micelle* | *micelle* | *vesicle* | *vesicle* |
| SANS | model | | *hollow sphere* | *sphere* | *sphere* | | |
| | $R$ | (nm) | 47.0 | 7.0 | 6.9 | | |
| | $\delta$ | | 14 | -- | -- | | |
| | $\sigma(R)$ *(Log-Normal)* | | 0.25 | 0.15 | 0.17 | | |
| | $\sigma(\delta)$ *(Log-Normal)* | | 0.17 | -- | -- | | |
| TEM | $R_{MET}$ | (nm) | 50 | 35 | 40 | | |
| | Preparation method | | *freeze-fracture* | *spray* | *freeze-fracture* | | |

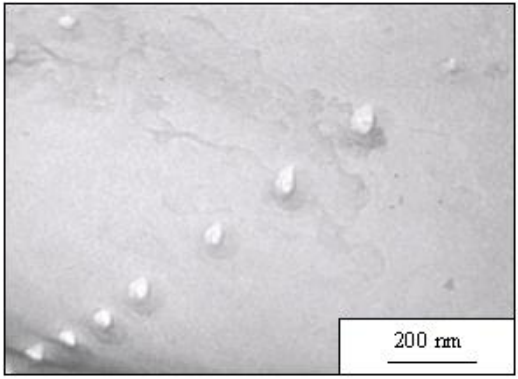

(a)

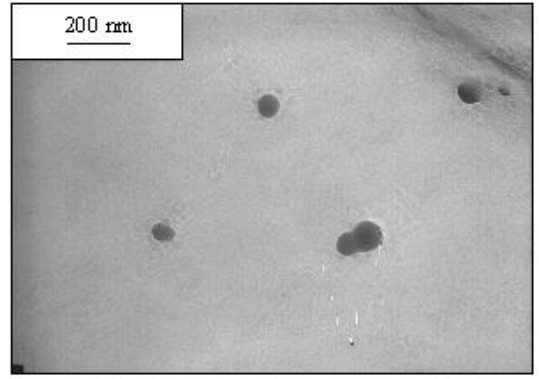

(b)

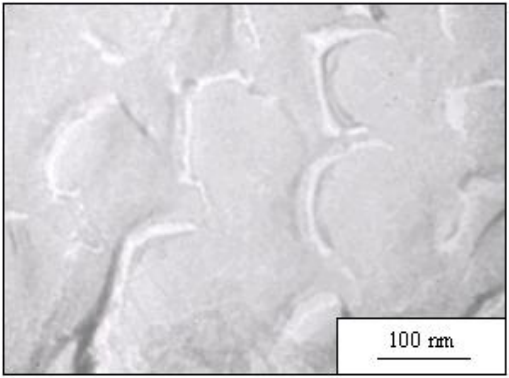

(c)

Figure 3: TEM analysis of copolymers (a) PB$_{48}$-*b*-PGA$_{145}$ (b) PB$_{48}$-*b*-PGA$_{114}$ and (c) PB$_{48}$-*b*-PGA$_{56}$. Samples (a) and (c) were prepared by freeze-fracture and (b) with the spray method.

In addition, small angle neutron scattering (SANS) experiments were performed in order to have more detailed structural data on the self-assembled micelles and vesicles.[23] All the samples were studied in pure D$_2$0 in order to increase the contrast between the copolymer and the solvent.

The scattering from N identical particles of spherical symmetry with a volume V$_p$ placed in a total volume V can be easily described as a function of the modulus of the scattering wave vector q:

$$I(q) = \frac{N}{V}(\Delta\rho)^2 V_p^2 P(q) S(q) \qquad (1)$$

where P(q) is the form factor of the particles, S(q) is the structure factor and $(\Delta\rho)^2$ the contrast, i.e. the difference in scattering length density between the particle and the solvent. In the case of very diluted solutions, the structure factor $S(q) \to 1$ and Eq. 1 becomes:

$$I(q) = \frac{N}{V}(\Delta\rho)^2 V_p^2 P(q) \qquad (2)$$

Typical scattering curves are shown in Figure 4 for $PB_{48}$-$b$-$PGA_{114}$ in $D_2O$ at pH=12 for concentrations ranging between 0.5 to 8% by weight. The scattered intensity is proportional to the concentration, except for low q values where the effects of repulsive interaction decrease the scattering intensity. Nevertheless, the structure factor has no influence on the form factor observed in a broad high q range. Here, we are dealing with a core-shell structure with a shell of PGA, i.e. a polyelectrolyte shell, the scattering of which is generally weak. On the contrary, the PB core is very dense and mainly contributes to the SANS signal. This phenomenon was developed theoretically by Shusharina et al.[24] for charged polymeric micelles: they evidenced that the compact micellar core is surrounded by a very extended corona of solvated hydrophilic segments. As a conclusion, the core-corona interface is sharp and it is then preferable to describe this interface as a border between the core and the solvent. Based on this assumption, we developed a model that take into account the intensity of an assembly of spherical PB domains (with a scattering length density $\rho_{PB} = 4.40 \times 10^9$ cm$^{-2}$) surrounded by heavy water (with a scattering length density $\rho_{D2O} = 6.38 \times 10^{10}$ cm$^{-2}$). The fit I(q) using a simple sphere model perfectly adjusts the experimental curve and gives a radius of the core $R_c$ of 7 nm. The value obtained for $PB_{48}$-$b$-$PGA_{145}$ micelles (Table 2) is comparable ($R_c$= 6.9 nm), then supporting the assumption that we are measuring the core size of PB only. Note that we accounted for the experimental resolution function in the fit.[25] Let us emphasize that the real polydispersity of the objects used to adjust the experimental curve is quite low: a log-normal distribution with $\sigma \sim 0.15$ has been used.

The aggregation number $N_{agg}$ in the micelles could be derived from the size of the core $R_c$ and the PB block density inside the core ($\rho_{PB}$=1.05 g.cm$^{-3}$ assuming a dense and solvent excluded core):

$$N_{agg} = 4\pi \rho_{PB} R_c^3 N_{Av} / 3\overline{M}_{PB} \qquad (3)$$

where $N_{Av}$ is the Avogadro number. We thus obtain $N_{agg}$ = 316 and 302 respectively for the copolymer micelles generated from PB$_{48}$-$b$-PGA$_{114}$ and PB$_{48}$-$b$-PGA$_{145}$. The surface area per chain $A_c = 4\pi R_c^2 / N_{agg}$ calculated for the two micellar systems gives the same value of 2nm$^2$. These results are in good agreement with the predictions of Zhulina and Birshtein[26] for hairy micelles (type IV) for which the core radius and the aggregation number are scaling only as a function of the composition of the block forming the core.

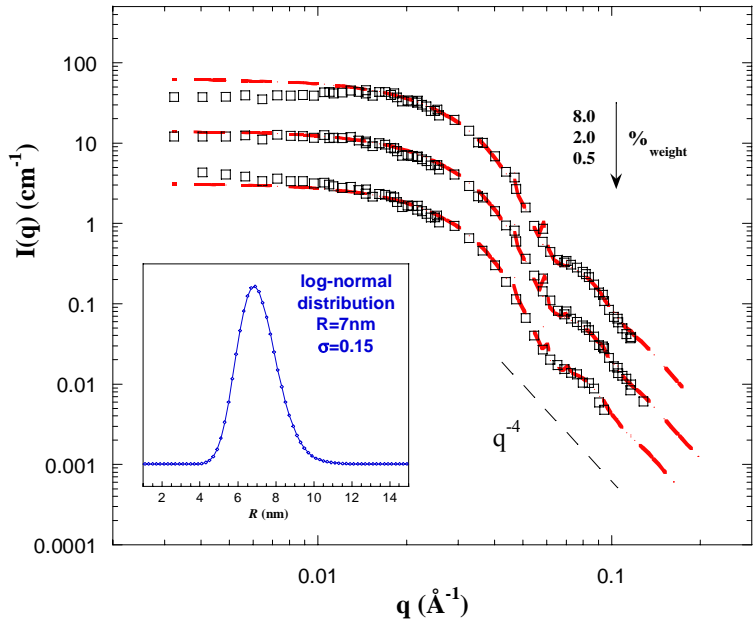

Figure 4: Small angle neutron scattering of PB$_{48}$-$b$-PGA$_{114}$ in D$_2$O at pH=12 for concentration ranging between 0.5 to 8% by weight. Broken lines are fitting curves by a simple sphere model. Insert: log-normal distribution of the micellar sizes.

The SANS data obtained for the copolymer PB$_{48}$-$b$-PGA$_{56}$ were different from the previous ones: the scattering intensity was much higher and there was no plateau at low q even for the very dilute solution. This means that the scattering objects are interacting and are larger than in the other copolymer solutions. This behaviour is in agreement with light scattering data. We have then assumed that the self-assembled structures obtained for this copolymer were no longer micelles but vesicles. From the slope of the representation $Ln(q^2I(q))$ versus q (Figure 5-insert) in the asymptotic Kratky-Porod approximation, we could estimate the thickness $\delta$ of the vesicles using the following scaling relation:

$$q^2 I(q) \propto \exp\left(-\frac{q^2 \delta^2}{12}\right) \quad q\delta < 1 \qquad (4)$$

We found $\delta = 13.8$nm. Figure 5 also shows the $q^2I(q)$ *versus* q representation of the SANS data which is appropriated for layer structures, since it enhances the oscillations of the form factor. Indeed, at large q values, the scattering intensity mainly depends on the membrane thickness and no longer on the global size of the vesicles. The fit is performed using the following "hollow sphere" form factor and leaving R and $\delta$ as free parameters:

$$P(q) = \frac{16\pi^2}{q^6}\left[\frac{\sin(qR_e) - qR_e\cos(qR_e) - \sin(qR_i) + qR_i\cos(qR_i)}{V_{R_e} - V_{R_i}}\right]^2 \qquad (5)$$

where $R_e$ and $R_i$ correspond respectively to the external and internal radii of the shell and $V_{Ri}$ ($V_{Re}$) is the volume of the sphere of radius $R_i$ ($R_e$). The fit leads to the following values for the radius (R = 47nm from the DLS analysis) and the membrane thickness ($\delta = 14$nm) that compares well with the one deduced from the fit to equation (2). Here again, the polydispersity of the membrane is low as the log-normal distribution used in the fit needed a $\sigma(\delta) = 0.17$. However, the close agreement between $R_G$ and $R_H$ seems surprising in the case of vesicles, as the thickness of the shell is such that $\delta/R$ is not negligible. We have checked by simple calculations that $R_G$ values are physically reasonable by assuming constant density profiles in the core and the corona. In the case of the micelles, in order to fulfill the observed values of $R_G$ and $R_H$ (DLS) and core size (SANS), the core-to-solvated-corona ratio of the scattering length densities is then ca. 14:1. Using this ratio and the same thickness (20nm) for the PGA-monolayer, the $R_G$ value of vesicles that should be measured by DLS is calculated to be 50.6 nm, very close to the observed value. It is unclear, however, why the hydrodynamic radius does not include the corona contribution in this case; this might be due to an even lower density.

Combining SLS, DLS, SANS and TEM, we have characterized the vesicles and micelles of the polypeptide diblock copolymers in water in basic conditions, i.e. when the polypeptide is charged and in a coil conformation.[27] Asymmetric diblocks with hydrophilic ratios around 70-75% form spherical micelles ($PB_{48}$-b-$PGA_{114}$ and $PB_{48}$-b-$PGA_{145}$). When the copolymer becomes symmetric ($PB_{48}$-b-$PGA_{56}$) the formation of vesicles is favoured. These data are in good agreement with the one observed for other amphiphilic block copolymers as described by Discher and Eisenberg.[28] A schematic representation of the self-assembled structures obtained in water is represented in Figure 6. When the hydrophilic content is too low, inverse structures are formed in organic solvents (THF and $CH_2Cl_2$) and the polypeptide block now is in the insoluble part of the aggregate. Even for asymmetric diblocks, vesicles could be obtained due to the rod-like conformation of the polypeptide that provides a flat interface, thus tying polymer conformation to supramolecular structure, as also observed by Deming.[3]

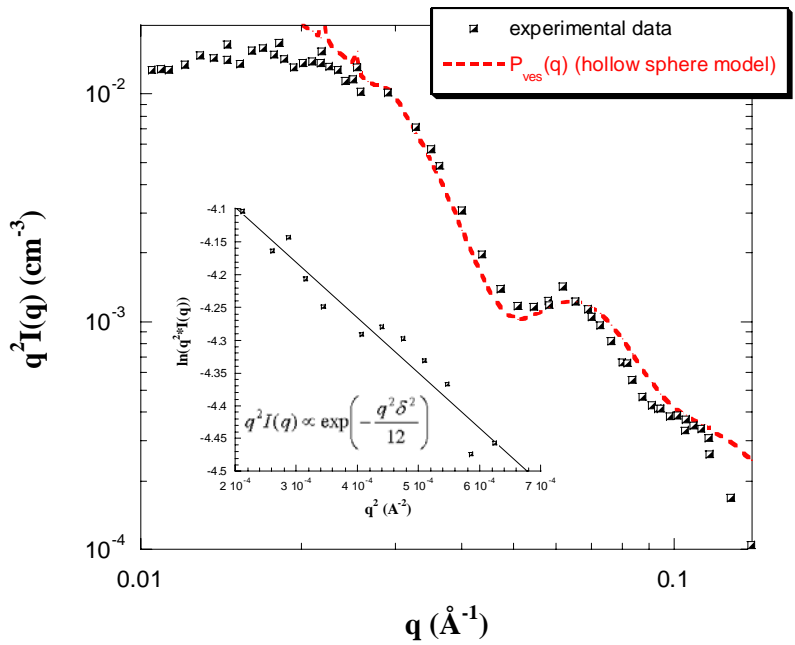

Figure 5: $q^2I(q)$ *versus* q representation of the small angle neutron scattering of $PB_{48}$-*b*-$PGA_{56}$ in $D_2O$ at pH=12 for a concentration c/w=2%. Broken line is a fitting curve by a simple hollow sphere model (eq. 5) with R=47nm ($\sigma$=0.25) and d=14nm ($\sigma$=0.17). Insert: measure of $\delta$ in the asymptotic regime (eq. 4).

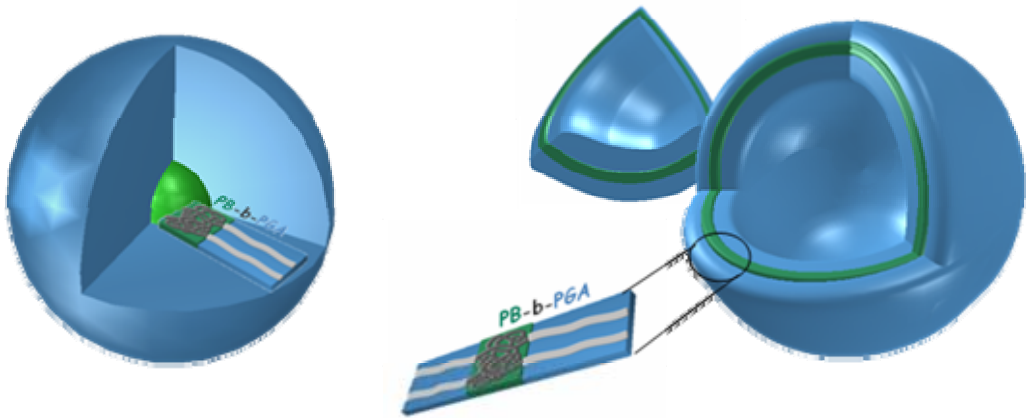

Figure 6: Schematic representation of micelles and vesicles obtained from the self-assembly of the polypeptide based diblock PB-*b*-PGA copolymers in water. The green part represent the polybutadiene and the blue part the poly(glutamic acid) segment.

### III-2- Stimuli-responsive effect to pH

Experiments described above have been performed in water at pH=12, meaning that the amino-acid side-chains on the glutamic acid residues were deprotonated, thus destabilizing the natural α-helical structure of PGA due to the electrostatic repulsion between charges. When decreasing the pH value, the PGA is expected to undergo a transition from coil to α-helical conformation whith the glutamic acid residues being protonated.[27] In order to study the helix-to-coil (or rod-coil) transition of PGA with pH inside the self-assembled structures, circular dicroïsm experiments have been carried out. The CD-UV spectrum of $PB_{48}$-*b*-$PGA_{56}$ vesicles (Figure 7) measured in water at pH=11.5 shows a typical double inflected curve with a small positive maximum at 218nm and a large minimum at 197nm, characteristic of a random coil conformation. When decreasing to pH=4.5, two negative minima at 221nm and 208nm and a positive maximum at 190nm could be detected, proving the formation of a α-helical conformation.[27] From the curve shown in figure 7, on can clearly conclude that the helix-to-coil transition occurs around pH=7.

The change in conformation induced by pH variation and its influence on the overall aggregation could be examined by DLS experiments (Figure 8). For all the block copolymer micelles and vesicles, no change in $R_H$ could be observed between pH=12 and 7. For pH values below 7, the hydrodynamic diameters decrease dramatically, exactly for the same conditions as the helix-to-coil transition. In fact, two effects occurred at the mean time: the decrease of the electrostatic repulsion forces between the amino-acid residues, and the change of conformation of the polypeptide chain, as a consequence of the protonation of the PGA. The first effect is classical and all kinds of polyelectrolytes have shown their ability to respond to pH variations.[6,7] However, when salt is added to these systems, they are generally not able to respond to pH variations as the electrostatic effect is screened. In order to decouple these two effects on our system, we have modified the ionic strength of solutions. Figure 9 represents the variation of the $R_H$ of micelles formed from copolymer $PB_{48}$-*b*-$PGA_{145}$ in water as a function of ionic strength and pH (the data are also presented in Table 3). This plot clearly evidences that even when the electrostatic interactions are completely screened (about 0.5 mol/l NaCl is needed), the micelles are still able to change their size with pH variations, yet with smaller amplitude. It is also interesting to mention that at acid pH, when the PGA block is not charged, there is no influence of the salt concentration on the measured $R_H$. However, in basic conditions (pH=12), the addition of salt results in a decrease of $R_H$ from 37nm to 28nm. The important consequence is that micelles are still able to respond in size to pH variations in high salt concentration conditions (as in biological media), proving that the helix-to-coil transition continues to occur. One has to mention at this stage that the pH-response is very fast (time for a DLS measurement), the size distribution of the micelles being always narrow.

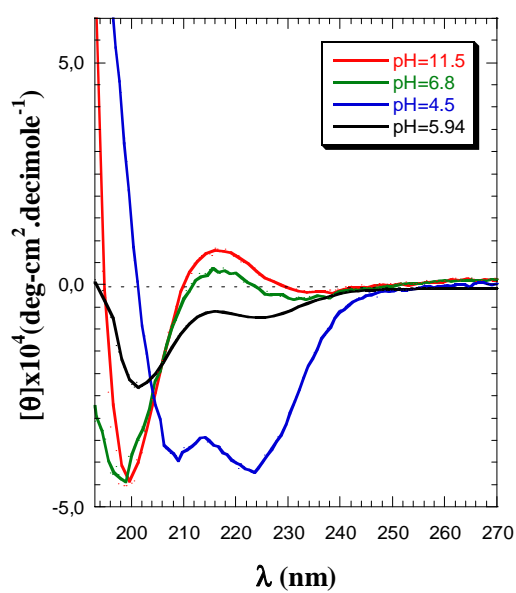

Figure 7: CD-UV Spectra of $PB_{48}$-*b*-$PGA_{56}$ vesicles measured in aqueous solutions for different pH values.

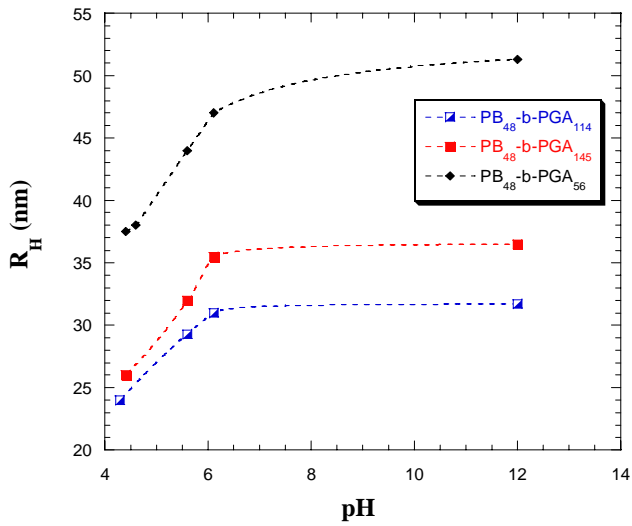 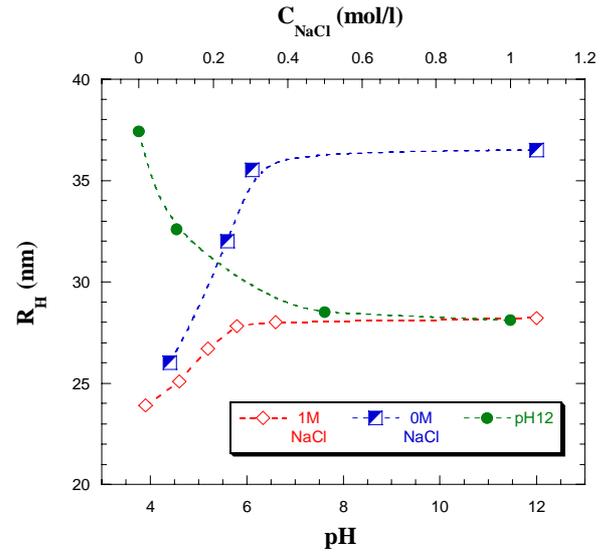

Figure 8: Hydrodynamic radius $R_H$ of PB-*b*-PGA micelles (■ PB$_{48}$-*b*-PGA$_{145}$, □ PB$_{48}$-*b*-PGA$_{114}$) and vesicles (♦ PB$_{48}$-*b*-PGA$_{56}$) as function of pH (no added salt).

Figure 9: Hydrodynamic radius $R_H$ of PB$_{48}$-*b*-PGA$_{145}$ micelles in water measured by DLS (90°) as a function of ionic strength (NaCl salt concentration) and pH.

However, one major question can be raised concerning the process of this responsive effect: is there any intermicellar exchange accompanied by a change in the aggregation number? In order to answer this question, SANS experiments have been carried out in acid and basic conditions on the same samples. As shown in Figure 10 for micelles of PB$_{48}$-*b*-PGA$_{114}$ copolymer, no real change could be determined when pH changes from 12 to 4. Fitting the core with a hard sphere model gives R=7.8nm at pH=4, instead of R=7nm at pH=12 (insert Figure 10 and Table 3). Similar change upon pH variation has been obtained for the copolymer PB$_{48}$-*b*-PGA$_{145}$. This apparent discrepancy in the micelle size has a simple explanation. Let us remember that for SANS, only the core of the micelles is visible. At low pH, the PGA chains adopt a α-helical conformation which is more compact and less hydrophilic than the extended coil conformation present for basic pH. As a consequence, the core appears larger due to the "collapse" of these PGA chains at the interface between the core and the shell. Moreover, the plateau reached for low q-values (even if the presence of a structure factor is present for basic conditions due to the electrostatic repulsion between the micelles), which is directly proportional to the molar mass of the micelles, is identical for the two pH values, indicating that the same object with the same molar mass and aggregation number is responsible for the scattered intensity.

Concerning the copolymer $PB_{48}$-*b*-$PGA_{56}$, which self assembles in vesicles at pH=12, the change upon pH variation is more pronounced than for the copolymer micelles. On Figure 11 showing $q^2I(q)$ *versus* q, we can see that the oscillations, characteristic of the membranes thickness, are now at larger q values in the case of pH=4. Upon adjusting the curves we found a decrease of the whole radius and the thickness of the membrane forming the vesicle when decreasing the pH (Table 3). As compared to micelles, the pH-induced changes in size that are observed for vesicles are larger and concern both the overall radius and the membrane thickness. The fact that this membrane thickness decreases with the pH probably reflects a more compact polybutadiene hydrophobic domain when PGA chains fold back to α-helical conformation with the decrease in pH from 12 to 4. Also, even if the low q-domain is difficult to adjust in basic conditions, due to the presence of a structure factor S(q) that is attributed to electrostatic repulsion of the vesicles, one can see that the plateau tends to the same value, reflecting that the same objects give the scattered intensity (Figure 11).

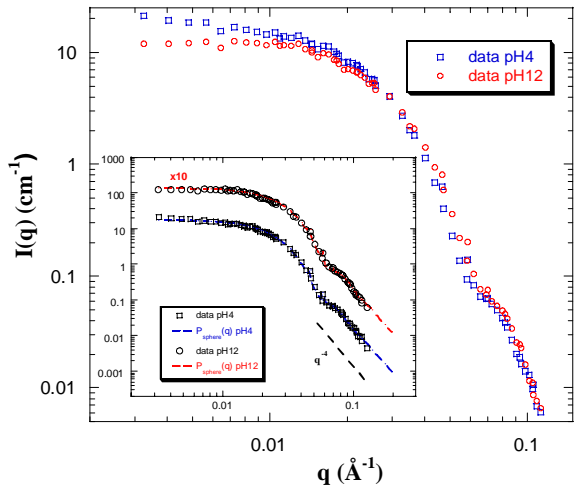 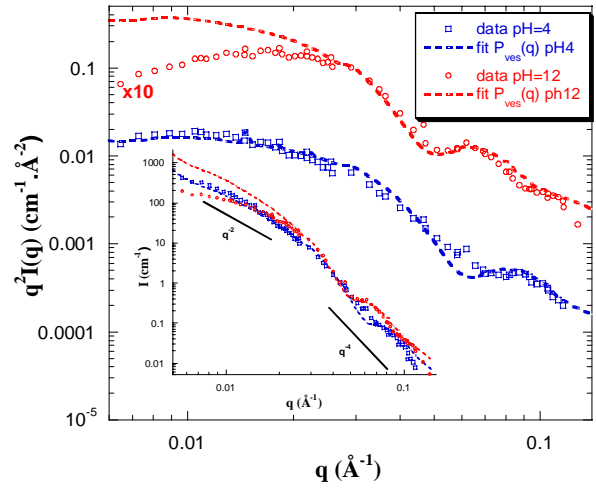

Figure 10: Small angle neutron scattering of $PB_{48}$-$b$-$PGA_{114}$ in $D_2O$ at pH=12 and 4 (c=2%). Insert: experimental data and fitting curves by a sphere model.

Figure 11: $q^2I(q)$ versus q representation of the small angle neutron scattering of $PB_{48}$-$b$-$PGA_{56}$ in $D_2O$ at pH=12 and 4 (c=2%). Broken lines are fitting curves by a hollow sphere model. Insert: I(q) versus q representation of the same samples.

**Table 3: DLS and SANS data obtained for the diblock copolymers PB-*b*-PGA as function of pH.**

| | $PB_x$-*b*-$PGA_y$ : | | $PB_{48}$-*b*-$PGA_{56}$ | $PB_{48}$-*b*-$PGA_{114}$ | $PB_{48}$-*b*-$PGA_{145}$ |
|---|---|---|---|---|---|
| **pH=12** | $R_{H,(0\ NaCl)}$ | (nm) | 53.5 | 32 | 37 |
| | $R_{H,(1M\ NaCl)}$ | | 47 | 27 | 28 |
| | model | | *hollow sphere* | *sphere* | *sphere* |
| | $R_{SANS}$ | (nm) | 47 | 7 | 6.9 |
| | $\delta$ | | 14 | -- | -- |
| | $\sigma(R)$ *(Log-normal)* | | 0.25 | 0.15 | 0.17 |
| | $\sigma(\delta)$ *(Log-normal)* | | 0.17 | -- | -- |
| **pH=4** | $R_{H,(0\ NaCl)}$ | (nm) | 37 | 24 | 26 |
| | $R_{H,(1M\ NaCl)}$ | | 37 | 24 | 26 |
| | model | | *hollow sphere* | *sphere* | *sphere* |
| | $R_{SANS}$ | (nm) | 37 | 7.8 | 7.5 |
| | $\delta$ | | 11 | -- | -- |
| | $\sigma(R)$ *(Log-normal)* | | 0.25 | 0.16 | 0.14 |
| | $\sigma(\delta)$ *(Log-normal)* | | 0.17 | -- | -- |

## IV- Conclusions

The polypeptide-based diblock copolymers described in this article form either well-defined self-assembled micelles or vesicles after direct dissolution in water, depending on their composition. The size of these structures could be reversibly manipulated as a function of environmental changes such as pH and ionic strength. Compared to other pH-responsive aggregates based on "classical" polyelectrolytes, these polypeptide-based systems present the ability to give a response in highly salted media as the chain conformational ordering can be controlled. This makes these systems suitable for biological applications: they provide significant advantages in the control of the structure and function of supramolecular self-assemblies. Moreover, polypeptides contain abundant chemical functionalities in the form of amino-acids that can be further used.

Polypeptide-based vesicles are certainly the most interesting structure and may offer many advantages compared to low molar mass lipid vesicles, in particular for applications in drug delivery.[28] Actually, in the past much emphasis and hope was placed on vesicles made of lipids or also called liposomes[29] for medical applications, but the latter were eventually found to be too unstable and their drug release properties too difficult to control. A vesicle obtained by self-assembly of a copolymer is expected to overcome some of these problems and thus allow the development of robust containers of either hydrophilic or hydrophobic species.

The development of macromolecular nanodevices that can be used within the living body implies that sensors detecting chemical signals -such as ions, enzymes or pH changes- and generating internal signals or appropriate responses be integrated in the macromolecular system. As evidenced here, the improvement brought about by polypeptides in such systems is related to the fact that the formed polymersomes can exhibit a variation in size as a function of the pH without any restriction due to ionic strength, thus making these systems suitable for *in vivo* applications. Moreover, the possibility to crosslinking these self-assembled nanoparticles makes them even more attractive as robust nano-containers with potentially long life time.[15] The use of a polypeptide thus seems to be crucial due to its unique response in physiological conditions and its biocompatibility, but also because it allows chemical modifications and grafting of either drugs or chemical bioreceptors, biosensors, etc…

**Acknowledgements:**

SL would like to thanks Pr. Harm-Anton Klok (EPFL, Lausanne, Switzerland) for previous collaborations and discussions.

**References:**

(1) (a) Ciferri A. *Supramolecular Polymers,* Marcel Dekker, New York, 2000. (b) Lehn, J.-M. *Supramolecular Chemistry - Concepts and Perspectives,* VCH Verlagsgesellschaft, Weinheim, 1995.

(2) (a) Klok, H.-A.; Lecommandoux, S. *Adv. Mater.* **2001**, *13,* 1217-1229. (b) Lee, M.; Cho, B.-K.; Zin, W.-C. *Chem. Rev.* **2001**, 101, 3869.

(3) Bellomo, E.; Wyrsta, M. D.; Pakstis, L.; Pochan, D. J.; Deming, T. J. *Nature Materials* **2004**, *3*, 244.

(4) Rodríguez-Hernández, J.; Lecommandoux, S. *J. Am. Chem. Soc., in press*.

(5) (a) Graff, A.; Sauer, P.; van Gelder, P.; Meier, M. *Proc. Natl. Acad. Sci. USA*, **2002**, *99*, 5604-5608. (b) Nardin, C.; Widmer, M.; Winterhalter, M.; Meier, M. *Eur. Phys. J. E*, **2001**, *4*, 403-410. (c) Chiu, D.T.; Wilson, C.F.; Karlson, A.; Danielson, A.; Lundqvist, A.; Stroemberg, F.; Ryttsen, M.; Davidson, F.; Nordholm, S.; Orwar, O.; Zare, R.N. *Chem. Phys.* **1999**, *247*, 133-139. (d) Kataoka, K.; Harada, A.; Nagasaki, Y. *Adv. Drug Delivery Rev.* **2001**, *47*, 113-131. (e) Lavasanifar, A.; Samuel, J.; Kwon, G. S. *Adv. Drug Delivery Rev.* **2002**, *54*, 169-190. (f) Kabanov, A. V.; Batrakova, E. V.; Alakhov, V. Y. *J. Controlled Release* **2002**, *82*, 189-212.

(6) Sauer, M.; Meier, W. *Chem. Commun.* **2001**, *1*, 55.

(7) Gohy, J.-F.; Willet, N.; Varshney, S.; Zhang, J.-X.; Jérôme, R. *Angew. Chem. Int. Ed.* **2001**, *40*, 3214.

(8) (a) Won, Y.-Y.; Davis, H.T.; Bates, F.S. *Science* **1999**, *283*, 960. (b) Stewart, S.; Liu, G. *Chem. Mater.* **1999**, *11,* 1048. (c) Thurmond, K.B. Huang, H.; Clark Jr, C.G.; Kowalewski, T.; Wooley, K.L. *Colloids and Surfaces B* **1999**, *16,* 45. (d) Armes, S. P. *J. Am. Chem. Soc.* **1998**, *20*, 12135.

(9) Rodríguez-Hernández, J.; Chécot, F.; Gnanou, Y.; Lecommandoux, S. *Prog. Pol. Sci., in press.*

(10) Chécot, F. ; Lecommandoux, S. ; Gnanou, Y. ; Klok, H.-A. *Angew. Chem. Int. Ed.* **2002**, *41,* 1339.

(11) Kukula, H.; Schlaad, H.; Antonietti, M.; Forster, S. *J. Am. Chem. Soc.* **2002**, *124*, 1658.

(12) Babin, J. ; Rodriguez-Hernandez, J. ; Lecommandoux, S. ; Klok, H.-A. ; Achard, M.-F. *Faraday Discussions*, **2004**, *128*, 179.

(13) Guangzhuo, R.; Mingxiao, D.; Chao, D.; Zhaohui, T.; Longhai, P.; Xuesi, C.; Xiabin, J. *Biomacromolecules* **2003,** *4, 1800.*

(14) Dong, C. -M. ; Sun, X.-S. ; Faucher, K. M.; Apkarian, R. P.; Chaikof, E. L. *Biomacromolecules* **2004**, *5*, 224.

(15) Chécot, F.; Lecommandoux, S.; Klok, H.-A.; Gnanou Y. *Eur. Phys. J. Part E* **2003**, *10*, 25.

(16) Ueda, K.; Hirao, A.; Nakahama, S. *Macromol.* **1990**, *23*, 939.

(17) Poché, D. S.; Moore, M. J.; Bowles, J. L. *Synth. Commun*. **1999**, *29*, 843.

(18) (a) Lecommandoux, S.; Achard, M.-F.; Langenwalter, J. F.; Klok, H.-A. *Macromolecules* **2001**, 34, 9100. (b) Klok, H.-A.; Langenwalter, J. F.; Lecommandoux, S. *Macromolecules* **2000**, 33, 7819.

(19) Cotton J.P. in *Neutrons, X-rays and Light scattering*, Lindner, P.; Zemb, Th. Eds.; Elsevier: North-Holland: Amsterdam, 1991; pp1-31.

(20) (a) Pedersen, J.S. *J. Phys. IV* **1993***, C8,3, 491 . (b)* Lairez, D. *J. Phys. IV* **1999**, *9*, 67.

(21) (a) Berne, B.J.; Pecora, R. *Dynamic Light Scattering*, Dover, 2nd Edition 2000. (b) Brown, W. *Dynamic Light Scattering* 1983, Clarendon Press, Oxford.

(22) Burchard, W. *Adv. Polym. Sci.* **1983***, 48,* 1-124.

(23) (a) Oberdisse, J. ; Porte, G. *Phys. Rev. E.* **1997***, 56*, 1965. (b) Oberdisse, J. ; Couve, C. ; Appell, J. ; Berret, J.F.; Ligoure, C.; Porte, G. *Langmuir* **1996**, *12*, 1212.

(24) Shusharina, N.P. ; Linse, P. ; Khokhlov, A.R. *Macromolecules* **2000**, *33*, 8488.

(25) Despert, G.; Oberdisse, J., *Langmuir* **2003**, 19, 7604-7610.

(26) Zhulina, E.B.; Birshtein, T. *Polym. Sci.* **1987**, *27*, 570.

(27) (a) Johnson, W.C.; Tonico, I. *J. Am. Chem. Soc.* **1972**, *94*, 4389. (b) Myer, Y.P. *Macromol*ecules **1969**, *2*, 624. (c) Adler, A.J.; Hoving, R.; Potter, J.; Wells, M.; Fasman, G.D. *J. Am. Chem. Soc.* **1968**, *90*, 4736.

(28) (a) Aranda-Espinoza, H.; Bermudez, H.; Bates, F. S.; Discher, D. E. *Phys. Rev. Lett.* **2001**, *87*, 208301. (b) Discher, D.E.; Eisenberg, A. *Science* **2002**, *297*, 967.

(29) D. D. Lasic, D. Papahadjopoulos, Eds., *Medical Applications of Liposomes* (Elsevier Science, Amsterdam, 1998).

**TOC graphic:**

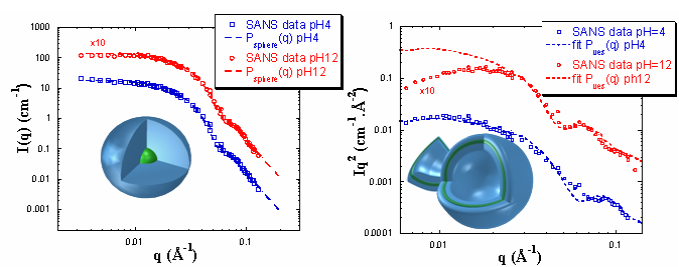

27